# Hydrodynamics of magnetic fluid droplets on superhydrophobic surfaces

**Nilamani Sahoo [a], Gargi Khurana [a], Devranjan Samanta [a,*,1], and Purbarun Dhar [a, b,*, 2]**

[a] Department of Mechanical Engineering, Indian Institute of Technology Ropar, Rupnagar–140001, India

[b] Department of Mechanical Engineering, Indian Institute of Technology Kharagpur, Kharagpur–721302, India

*Corresponding authors:

[1] E-mail: devranjan.samanta@iitrpr.ac.in

[1] Tel: +91-1881-24-2109

[2] E-mail: purbarun@mech.iitkgp.ac.in;

[2] Tel: +91-3222-28-2938

## Abstract

The study reports the aspects of post-impact hydrodynamics of ferrofluid droplets on superhydrophobic (SH) surfaces in the presence of a horizontal magnetic field. A wide gamut of dynamics was observed by varying the impact Weber number (We), the Hartmann number (Ha) and the magnetic field strength (manifested through the magnetic Bond number ($Bo_m$)). For a fixed We~60, we observed that at moderately low $Bo_m$~300, droplet rebound off the SH surface is suppressed. The noted We is chosen to observe various impact outcomes and to reveal the consequent ferrohydrodynamic mechanisms. We also show that ferrohydrodynamic interactions leads to asymmetric spreading; and the droplet spreads preferentially in a direction orthogonal to the magnetic field lines. We show analytically that during the retraction regime, the kinetic energy of the droplet is distributed unequally in the transverse and longitudinal directions due to the Lorentz force. This ultimately leads to suppression of droplet rebound. We study the role of $Bo_m$ at fixed We~60, and observed that the liquid lamella becomes unstable at the onset of retraction phase, through nucleation of holes, their proliferation and rupture after reaching a critical thickness only on SH surfaces, but is absent on hydrophilic surfaces. We propose an analytical model to predict the onset of instability at a critical $Bo_m$. The analytical model shows that the critical $Bo_m$ is a function of the impact We, and the critical $Bo_m$ decreases with increasing We. We illustrate a phase map encompassing all the post-impact ferrohydrodynamic phenomena on SH surfaces for a wide range of We and $Bo_m$.

## 1. Introduction

Collisional and post-impact hydrodynamics of droplets on solid surfaces has been an active area of research, revealing various physical phenomena like rebound, partial rebound, splashing, fragmentation and deposition [1, 2]. Understanding droplet impact dynamics is relevant for different applications like inkjet printing, spray cooling, spray coating, or efficient deposition of pesticides on vegetation, etc. [1, 2]. Although research on impact dynamics of fluid droplets has led to an enormous and rich body of literature, droplet impact dynamics research with utilitarian implications are often rare. To this end, one important study is the droplet impact hydrodynamics of ferrofluids under the influence of magnetic field. Understanding post-impact dynamics of ferrofluids droplets in magnetic field ambience will prove useful in improving magnetic 3-D printing performance wherein droplet spreading dynamics promotes shape distortions in metallic droplets during laser-induced curing [3, 4]. Another important utility of magnetic fluid impact dynamic will be towards optimization of liquid metal droplet deposition during welding or soldering. Magnetic fluid droplet manipulation is also of prime utility in several microfluidic devices and systems.

Ahmed et al. [5] studied droplet impact dynamics in horizontal and vertical magnetic field environments. The deviation of the pre-impact droplet shape from spherical to ellipsoids due to the magnetic field and subsequent spreading dynamics were discussed. Subsequently, the same research group reported an analytical model to predict the spreading dynamics under different governing parameters, like Weber number ($We = \frac{\rho V_o^2 D_o}{\sigma_{lv}}$ defined as the ratio of inertial force to surface tension, where $V_o$ is the impact velocity, $D_o$ is the pre-impact diameter of the droplet and $\rho$ and $\sigma_{lv}$ represent the density and surface tension of the ferrofluids, respectively), the magnetic Bond number ($Bo_m = \frac{B^2 D_o}{\mu_o \sigma_{lv}}$ defined as the ratio of magnetic force to surface tension, where $B$ is the magnetic flux density, and $\mu_o$ is the magnetic permeability of free space) and the impact Reynolds number (Re). Yang et al. [6] reported the elliptical spreading of liquid metal (GaInSn alloy) droplets on glass surfaces under the influence of a horizontal magnetic field. Using simulations, they deduced the distribution of induced magnetic field lines and correlated the non-uniform radial distribution of the Lorentz force to the elliptical spreading behavior.

Duvivier et al. [7] used a magnet below a horizontal superhydrophobic substrate to modify the gravitational effects during droplet impact, and explored the different dynamics under different We, $Bo_m$ and Re. Sudo et al. [8] studied the influence of the magnetic field strength on the maximum spreading diameter and the splashing dynamics. Spikes (albeit not similar to the Rosenswieg instability) were observed in the case of magnetic field perpendicular to the substrate, but not for the case of field parallel to the substrate. Rahimi & Weihs [9] studied the droplet impact dynamics of magnetorheological (MR) fluids and showed that both the spreading radius and spreading velocity is decreased in presence of vertical magnetic field. Another study on MR fluids [10] reported the fragmentation of droplets initiated by surface instability under the influence of strong vertical magnetic field. Zhou and Jing [11] studied the effect of impact velocity, magnetic field intensity and magnetic properties of the fluid on the spreading diameter and height of post impact ferrofluid droplets.

The present article reports the droplet impact hydrodynamics and regimes of different ferrofluid droplets under varying horizontal magnetic field intensities. For different governing We and $Bo_m$, we have observed various post-impact dynamics, like rebound, suppression of rebound, fragmentation of the lamella during retraction, and initiation of holes within the lamella leading to rupture of the liquid film. A brief review on suppression of droplet rebound and fragmentation of the liquid lamella during retraction in various hydrodynamic scenarios has been presented subsequently. Antonini et al. [12] reported that water droplets can stick to superhydrophobic (SH) surfaces via the Cassie-to-Wenzel transition (CWT), when the receding contact angle becomes higher than $100^o$. Mao et al. [13] proposed the dependence of rebound behavior on the spreading factor and the static contact angle. Non-Newtonian droplets were observed to exhibit rebound suppression on SH surfaces due to generation of normal stress and subsequent slowdown of the contact line velocity during onset of retraction [14-16]. The rebound suppression of such droplets was shown to be dependent on impact velocity, polymer concentration, and a critical Weissenberg number based on shear rate at the moving contact line during retraction [17].

Yun and Lim [18] investigated the effects of the shape eccentricity of the pre-impact droplet (induced by electric field) and the Weber number on rebound suppression phenomenon. Instead of uniform radial retraction, the droplets alternately expand and contract along the horizontal axes, thereby reducing the kinetic energy otherwise available at retraction, which leads to rebound suppression. Lee and Kim [19] showed that the droplet rebound mechanism can also be modulated by altering the motion of the solid target along the vertical direction. Recently, Yun [20] reported the impact dynamics outcomes of truncated spherical drops upon solid surfaces, as function of the truncation depth, surface wettability, and impact velocity. In addition to rebound suppression, our observations also reveal that at high $Bo_m$, holes are nucleated within the retracting film or lamella, ultimately leading to its rupture. Previous studies have shown that with increase of impact We, the liquid lamella undergoes similar internal rupture [21]. Biance et al. [22] studied the rupture of the expanding lamella through nucleation of holes and subsequent growth during Leidenfrost impact of droplets.

To the best of knowledge, systematic understanding of the hydrodynamics of ferrofluid droplets post-impact on SH surfaces in magnetic field environment is yet to be reported. Additionally, the role of magnetic field on rebound suppression phenomena, the kinetics of post-impact spreading and fragmentation of liquid lamella are yet to be reported.

In this article, we present experimental evidence of asymmetric spreading regimes, rebound suppression of ferrofluid droplets on a SH surface in presence of horizontal magnetic field. We explain the ferrohydrodynamics of the droplets by appealing to the principles of fluid dynamics, interfacial behavior and Lorentz forces, and their interactions. We also propose an analytical formalism to explain the dependence of the $Bo_m$ on the inhibition of rebound for a fixed impact velocity (or We). Subsequently, we show that at higher $Bo_m$, the post-impact droplets undergo orthogonal fragmentation, which is otherwise absent for impacts without magnetic stimulus. The nucleation of internal holes in the liquid film after impact, ultimately leading to proliferation, to fragmentation and rupturing of the droplet has been modeled theoretically. We have found excellent agreements between the experimental and theoretical critical $Bo_m$ at which such instability is triggered. In addition, we also show how the $Bo_m$ and the surface superhydrophobicity interplay towards governing the paradigms of impact hydrodynamics in presence of magnetic field.

## 2. Materials and methodologies

A schematic of the experimental setup has been illustrated in fig. 1. The experiments were conducted using a droplet dispenser with a digitized control unit, with a volumetric accuracy of ± 0.1μl. The ferrofluids were dispensed using a precision glass micro-syringe with a 22-gauge needle. The magnetic field was generated between the poles of an electromagnet (Holmarc Opto-Mechatronics Ltd., India). A DC power source (Polytonic Corp., India) was used to supply the electromagnetic coils. A GaAs sensor-based Hall effect Gauss meter (Holmarc Opto-Mechatronics Ltd., India) was used to calibrate the magnetic field strength at the center of the poles by varying the current input. The needle was carefully positioned for the droplets to fall at the center of the pole spaces. The gap between the poles was maintained 14 mm throughout the study. A high-speed camera (Photron, UK) mounted with a G-type AF-S macro lens of constant focal length 105 mm (Nikkor, Nikon) was used to capture the impact hydrodynamics. The images were recorded at 3600 frames per second at 1024x1024 pixels resolution.

Spray coated (Neverwet Ultra Everdry, USA) superhydrophobic glass slides were used as target surfaces. The surfaces were prepared following previously reported protocol [23]. The SH slides were placed between the poles of the electromagnet horizontally to ensure droplet impact at the center of the poles. Water based stable ferrofluids (made with 5-8 wt. % $Fe_2O_3$ and $Fe_3O_4$ nanoparticles) were used as test fluids. The colloidal solutions were synthesized in-house using ultrasonic disruption and were noted to be stable for time periods largely exceeding the experimental timescales. Katiyar et al [24] reported the characterization and preparation of such ferrofluids which are stable over a range of magnetic field strengths (0-1.2 T) and similar method has been adapted in the present study. The viscosity of the different ferrofluids (in presence and absence of magnetic field) was measured by a rotational rheometer (Anton Paar) with parallel plate geometry, attached to a magnetorheology module.

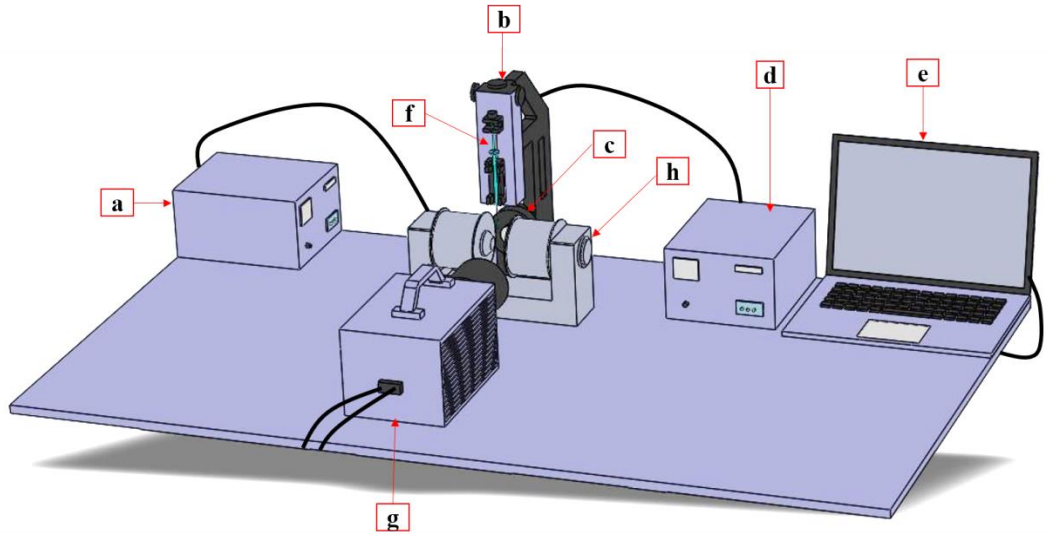

**Fig.1** Schematic of experimental setup: (a) Electro magnet power controller and supply (b) Droplet dispensing mechanism (DDM) unit (c) Backlight (d) DDM and backlight illumination controller (e) Computer for data acquisition and camera control (f) Micro syringe (g) High speed camera (h) Programmable electromagnet unit

Surface tension of the liquids was determined by the pendant drop method and image analysis. The contact angles were determined from image analysis using ImageJ software. The equilibrium contact angles ($\theta_e$) of the ferrofluids on glass, Teflon and SH surface are measured as ~37°, ~87°, and ~145.6° respectively. The ferrofluids are chosen for a range of impact Hartmann numbers ($Ha = \dfrac{\rho m B D_o}{2 V_o \eta}$), which represents the ratio of magnetic force to viscous force [25], where $\eta$ is the viscosity of the ferrofluids and m is the magnetic moment of the droplet to observe various dynamical outcomes of post impact droplets. The fluid properties like density, viscosity and surface tension and size of the ferrofluids at 25°C are given in Table 1. The variation in properties and size of ferrofluid droplets generated are within ±5%.

**Table 1:** Associated properties of the ferrofluids

| Parameter | $Fe_3O_4$ ferrofluid | $Fe_2O_3$ ferrofluid |
|---|---|---|
| Density (kg/m$^3$) | 1060.0 | 1061.5 |
| Viscosity (Pa-s) | 0.07754 | 0.054 |
| Surface tension (N/m) | 0.075 | 0.075 |
| Hartmann number, Ha | 0–8.0 | 0–13 |
| Diameter (mm) | 2.81 | 2.8 |

# 3. Results and discussions

We have categorized the droplet impact ferrohydrodynamics into two categories based on the magnetic Bond number $Bo_m$ ( $Bo_m = \dfrac{B^2 D_o}{\mu_o \sigma_{lv}}$ ). The first regime is up to $Bo_m \sim 300$, at which droplet rebound is arrested. The second regime is at higher $Bo_m \geq 300$; where fragmentation of the lamella during retraction is observed. At We ~40 or less, experiments cannot be conducted beyond $Bo_m > 10$, as the needle is present within the influence of the magnetic field, which interferes with the droplet generation process at the needle. In these conditions (We<40), the magnetic field prevents the free fall of the droplet from the needle. In this case the magnet attracts the droplet towards the poles, thereby hampering the vertical free fall on the target surface. Therefore, we vary the impact velocities (We ~5-125) to observe various impact outcomes in absence of magnetic field whereas the experiments (We≥40) have conducted for a broad range of $Bo_m$ (from ~0–5500) on SH surfaces.

**(i) Droplet rebound at low $Bo_m \leq 300$ and We~60** Droplet dynamics after impact on SH surfaces in magnetic field influence for $Bo_m \leq 300$ and We~60 are summarized and illustrated in fig.2. Fig. 2a describes the temporal evolution of the droplets after impact on SH surfaces at different $Bo_m$ (≤300) for a fixed We ~ 60. At $Bo_m=0$, the droplet exhibits complete rebound, with the formation of a secondary droplet. As $Bo_m$ is increased up to ~300, droplet rebound is suppressed. The suppression dynamics is quantified using the elongation factor. Fig. 2b shows the evolution of the elongation factor (β) for different magnetic fields. The β is expressed as the ratio of the perpendicular height from the point of contact of the pre-impact droplet at the solid surface to the top of droplet (L) to the initial droplet diameter ($D_o$). As can be seen from fig.2b, for $Bo_m=0$, the elongation factor increases up to ~3 after the retraction phase, and followed by a sharp decrease near the peak due to secondary droplet pinch-off. At increasing $Bo_m$ and approaching ~300, the elongation factor is largely reduced due to droplet rebound suppression, and assumes a steady value of ~1.

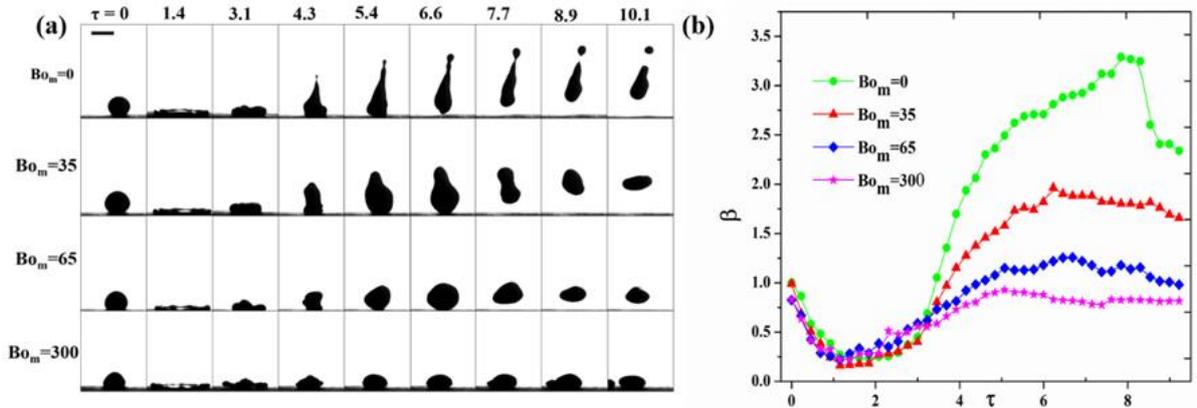

**Fig.2** (a) side view images of post impact droplets on superhydrophobic surfaces (SH) for We=60 and different $Bo_m$. The scale bar corresponds to 2.81 mm. τ represents the non-dimensional time of post impact droplet. The non-dimensional time $(\tau=V_o t/D_o)$ is defined as the ratio of the impact velocity ($V_o$) times elapsed time (t) to the initial droplet diameter ($D_o$). (b) Variation of elongation factor ($\beta$) with non-dimensional time ($\tau$).

The magnetic field is observed to lead to shape distortions during the impact phase, which leads to asymmetric spreading regimes. Similar behavior in droplets was noted using electric body force shape distortions [26]. The side views (fig. 2a) were further supplemented with top views (fig. 3a). It is clearly evident that with increase in $Bo_m$, the droplets deviate from the usual radial contraction ($Bo_m=0$, fig. 3a top row) during retraction phase. The droplets also undergo asymmetric spreading, with more spreading propensity orthogonal to the direction of the magnetic field ($Bo_m=300$, fig. 3a bottom row). In case of the usual uniform radial spreading, after attaining the maximum spread state, the kinetic energy is primarily transferred along the direction perpendicular to the impacting surface ($Bo_m=0$, fig 3 (a)), which leads to rebound at the end of retraction. The deviation from the radially symmetric shape at higher $Bo_m$ is quantified using the ratio of post impact spreading radius in two mutually orthogonal axes (transverse and longitudinal) as $\xi = \dfrac{R_{trans}}{R_{long}}$. The orthogonal spreading ratio $\xi$ is observed to increase with the increase in magnetic field, implying higher deviation from radially symmetric spreading (fig. 3b). At higher $Bo_m$, instead of uniform radial retraction, the diversion of spreading kinetic energy in the transverse direction reduces the amount of retraction kinetic energy in direction normal to the surface, leading to rebound suppression. In Fig. 3b, reduction in the $\xi$ is noted in the initial regime at high $Bo_m$, which is caused by morphing of the spherical droplet to elliptical at the point of impact due to high magnetic field strength. Similar dynamics has also been noted in literature [27, 28].

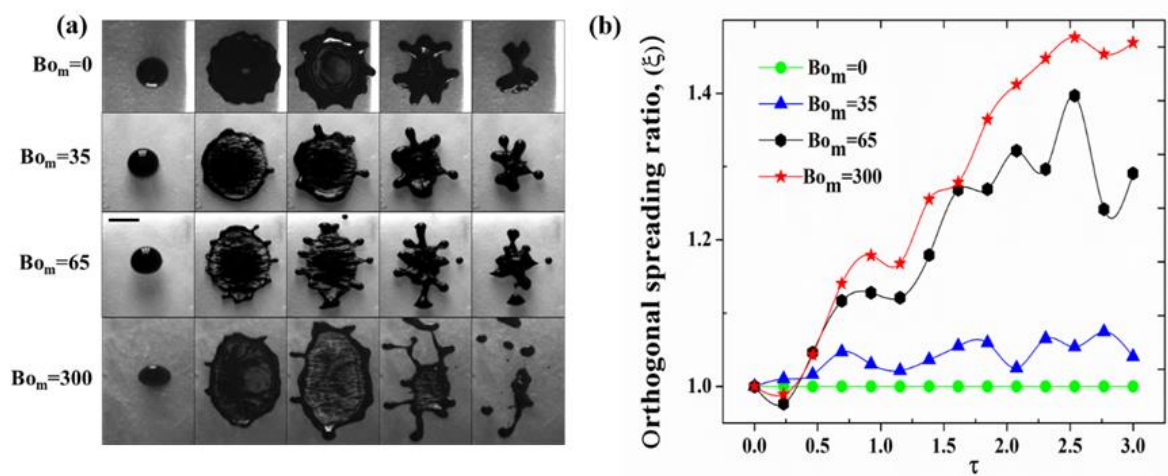

**Fig.3** (a) Top view images of post impact droplets on SH surfaces for We=60 and different $Bo_m$. The spreading of the droplet occurs from left to the right for a fixed $Bo_m$. Time evolves along each row at 2.16 ms intervals. The scale bar is 4 mm. In each case the magnetic field acts horizontally across the droplet. (b) Variation of the orthogonal spreading ratio with the non-dimensional time.

We have developed an analytical formalism to explore the effect of magnetic field modulated retraction velocity on the rebound suppression. During asymmetric spreading in the direction orthogonal to the field lines, the droplet attains a thin liquid lamella structure of thickness (h) and radius (R) (measured along the direction of the magnetic field) in the presence of horizontal magnetic field. Then the spread out droplet rapidly retracts forming a rim that collects the liquid present in the lamella (Fig. 3a, $Bo_m$=300, 3$^{rd}$ image). Along the lines of a previous report [29], the dynamics of the droplet can be determined by force balance between the surface tension of the liquid lamella, the magnetic force, and the inertia of the rim. The analytical expression for the retraction velocity is derived along the lines of an existing framework [29]. We can write the force conservation for the liquid rim in the direction perpendicular to the magnetic field direction as:

$$\frac{d}{dt}\left(m_r \frac{dr}{dt}\right) = F_c + F_m \quad (1)$$

Where $m_r$ is the mass of the rim, r is the instantaneous radius of the rim, $F_c$ is the capillary tension acting on along the rim, and $F_m$ is the magnetic force acting on the droplet due to presence of the horizontal magnetic field. The capillary force can be expressed as [29]:

$$F_c = 2\pi R \sigma_{lv} (1-\cos\theta_r) \quad (2)$$

where, $\theta_r$ is the receding contact angle. Clearly, when we consider the R along the direction orthogonal to the magnetic field, there will be an overestimation of capillary force (see fig. 3a) as the droplet spreads more in that direction compared to the direction of field. However, the magnitude of the capillary force is much smaller than the magnetic force, so the overestimation of the capillary force will be not significant with the increase of $Bo_m$ (see fig. 4). Similarly, the magnetic force acting on the ferrofluid droplets [5] is expressed as:

$$F_m = \frac{\pi}{4} \frac{B^2 \chi D_o^2}{\mu_o (1+\chi N)} \quad (3)$$

where $\chi$ is the magnetic susceptibility of the ferrofluid and N is the demagnetizing factor. The magnetic force has been deduced by considering the derivative of the magnetic energy of the droplet with respect to the initial droplet diameter [5]. The values of N and $\chi$ for the ferrofluids are obtained from literature [5, 30].

Further, at the onset of retraction phase the inertia of the rim associated with its acceleration are neglected against the capillarity [29, 31]. We, therefore, express the inertia of the receding rim as:

$$F_{rim} = \dot{m}_r \frac{dr}{dt} \quad (4)$$

where $\dot{m}_r = \rho(2\pi Rh)V_{ret}$ and $\dfrac{dr}{dt} = V_{ret}$. $V_{ret}$ is the retraction velocity.

Finally, the inertia force can be expressed as:

$$F_{rim} = \rho(2\pi Rh)V_{ret}^2 \quad (5)$$

Substituting Eqs (2), (3) and (5) in Eq (1), the final momentum equation can be obtained as

$$2\pi Rh\rho\, V_{ret}^2 = 2\pi R\sigma_{lv}(1-\cos\theta_r) + \frac{\pi}{4}\frac{B^2\chi D_o^2}{\mu_o(1+\chi N)} \quad (6)$$

At the moment the droplet attains the maximum spreading diameter ($R_{max}$), we can estimate the value of thickness of the droplet from conservation of volume as

$$h_m = \frac{4}{3}\frac{R_o^3}{R_{max}^2} \quad (7)$$

So, the final retraction velocity is obtained using Eq. (7) as

$$V_{ret} = \left(\frac{\sigma_{lv}}{h_m\rho}(1-\cos\theta_r) + \frac{\sigma_{lv}}{4h_m\rho}\frac{\chi}{\psi_m(1+\chi N)}Bo_m\right)^{\frac{1}{2}} \quad (8)$$

where $\psi_m$ is the maximum spreading factor which is defined as the ratio of the diameter of the droplet at maximum spread state to the initial pre-impact diameter of the droplet $\left(\dfrac{D_{max}}{D_0}\right)$.

In a similar method, we determine the ratio of the retraction velocity along the longitudinal and transverse directions, just after the post-impact droplet realizes the maximum spreading diameter as

$$\frac{V_{ret,\,long}^2}{V_{ret,\,trans}^2} \approx \frac{\dfrac{\sigma_{lv}}{h_{m,long}\rho}(1-\cos\theta_r) + \dfrac{\sigma_{lv}}{4h_{m,long}\rho}\dfrac{\chi}{\psi_{m,long}(1+\chi N)}Bo_m}{\dfrac{\sigma_{lv}}{h_{m,trans}\rho}(1-\cos\theta_r) + \dfrac{\sigma_{lv}}{4h_{m,trans}\rho}\dfrac{\chi}{\psi_{m,trans}(1+\chi N)}Bo_m} \quad (9)$$

The receding contact angle ($\theta_r$) and the thickness ($h_m$) are considered at the instant of maximum spread. With increase in the $Bo_m$, the second component of the numerator as well as denominator on the R.H.S. of equation (7) becomes dominant over the interfacial energy component $\left(\dfrac{\sigma_{lv}}{h_m\rho}(1-\cos\theta_r)\right)$ (see fig. 4). Hence at high $Bo_m$, the R.H.S. of equation 9 reduces to the ratio of the magnetic forces in the longitudinal (orthogonal) and the transverse direction. The eqn. 9 can be further simplified to

$$\frac{V_{ret,long}^2}{V_{ret,trans}^2} \approx C \frac{\psi_{m,trans}}{\psi_{m,long}} \quad (10)$$

Now, C is a constant parameter that is the ratio of $h_{m,trans}$ and $h_{m,long}$. The value of C depends on the orthogonal spreading ratio $\zeta$ and falls in the range of $0<C\leq1$. The limiting case $C \sim O(1)$, for uniform radial spreading in absence of magnetic field. From equation 10, it is evident that the retraction velocity is inversely proportional to square root of maximum spreading factor, which is caused by the presence of the magnetic force. Since $\psi_{m,long} < \psi_{m,trans}$, the droplet retracts much faster in the longitudinal direction. This, in turn, produces asymmetric retraction velocity along the mutual orthogonal axes, ultimately leading to elongation of the drop in transverse direction. In addition, the reduction of retraction velocity in the orthogonal direction leads to arrest of the rebound of the droplet at end of retraction.

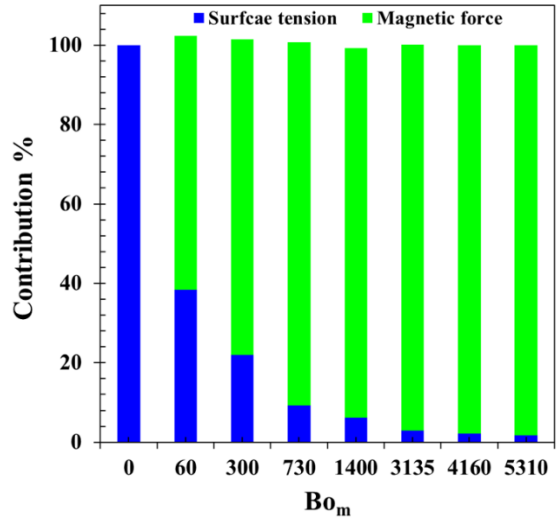

**Fig.4** Impact of the $Bo_m$ on the surface tension and magnetic force on the droplet. The surface tension and magnetic force components are expressed as $\left(\frac{\sigma_{lv}}{h_m\rho}(1-\cos\theta_r)\right)$ and $\left(\frac{\sigma_{lv}}{8h_m\rho}\frac{\chi}{\psi_m(1+\chi N)}Bo_m\right)$ respectively.

**(ii) Impact ferrohydrodynamics at $Bo_m >300$ and We=60** With the increase in magnetic field strength ($Bo_m\sim5310$), fragmentation of the liquid lamella during the retraction phase is observed (Fig. 5 (a) & (c), columns (ii)-(iv)). As soon as the orthogonal spreading ratio attains a value of approximately $\xi\sim1.5$, the droplet is noted to contract longitudinally and expand in the transverse direction (Fig.5 (b) and (d)). Until now our experimental studies were restricted to low Hartman number Ha ($\leq8$). Subsequently, with different type of magnetic particles ($Fe_3O_4$), we increased the Ha upto 13. At a higher value of Ha and at $\tau=3.2-3.8$ (Fig. 5 (b) $Bo_m=1400$), the rim merged along the transverse direction during retraction phase, to form a filamentous structure with several ejected daughter droplets. Since the impact We remains unchanged, the morphing of impact dynamics is caused by the ferrohydrodynamic forces due to the magnetic fluid.

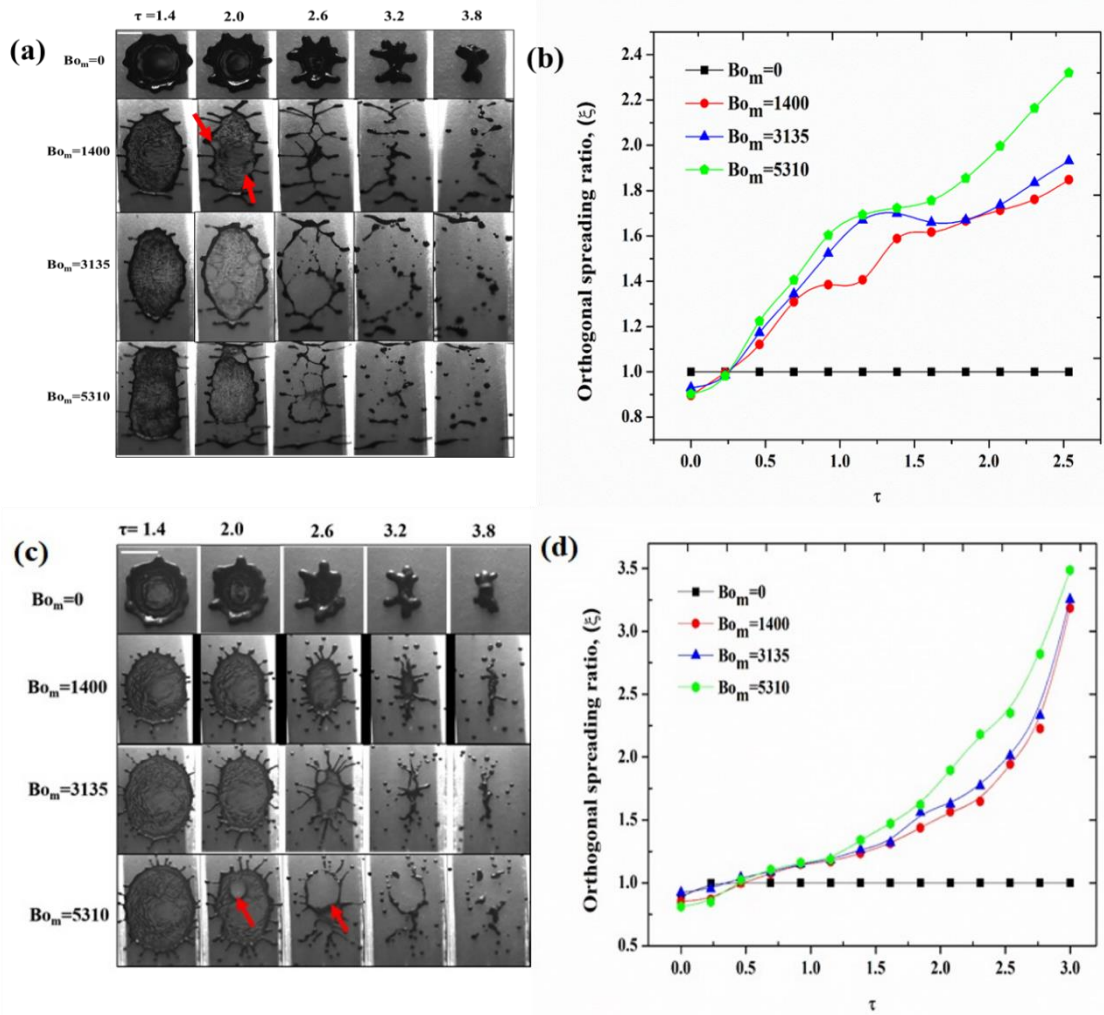

**Fig.5** (a) and (c) respectively show the top view images of low and high Hartmann number ferrofluid droplets at We=60. (a) Represents up to Ha≤8.0 and (c) represents up to Ha≤13. A clear effect of the Ha is observed when comparing the $Bo_m$=1400 cases in (a) and (c). Similarly, (b) and (d) show the variation of orthogonal spreading with time at the same We. The orthogonal spreading is largely prominent at high $Bo_m$. The arrows show the nucleation of the hole within the lamella and its growth. The scale bar is 6.9 mm.

Subsequently, we propose a mathematical analysis to highlight the role of the magnetic force on orthogonal spreading characteristics (see fig. 5) at different $Bo_m$. The magnetic force component along the longitudinal axis [5] is expressible as:

$$F_L = \frac{\pi}{4} \frac{B^2 \chi D_o^2}{\mu_o (1+\chi N)} \qquad (11)$$

The magnetic torque acting on the ferrofluid drop in presence of the horizontal magnetic field is thereby

$$\tau = \mu_o m \times H \qquad (12)$$

Where, H is the magnetic field intensity and its magnitude is equal to $\frac{B}{\mu_o(1+\chi N)}$. m is the magnetic moment. So, the magnetic force along the transverse direction can be written as

$$F_T = \frac{mB}{(1+\chi N)D_o} \qquad (13)$$

The magnetic forces in the longitudinal direction (eqn. 11) and in the transverse direction (eqn. 13) are directly and inversely proportional to the spreading diameter, respectively. Therefore, as the droplet tries to spread longitudinally, the magnetic force grows in magnitude and opposes the hydrodynamic motion. On the contrary, the magnetic force in the transverse direction decays in magnitude, thereby promoting the droplet spreading in the transverse direction. This scaling analysis thereby explains the observed asymmetric droplet spreading, which is essentially caused by interactions between the magnetic and the hydrodynamic forces during spreading. Based on eqns. 11 and 13, it is evident from fig. 6 that with increase in the $Bo_m$, the non-dimensional magnetic force in the transverse direction is an order of magnitude higher than its longitudinal counterpart.

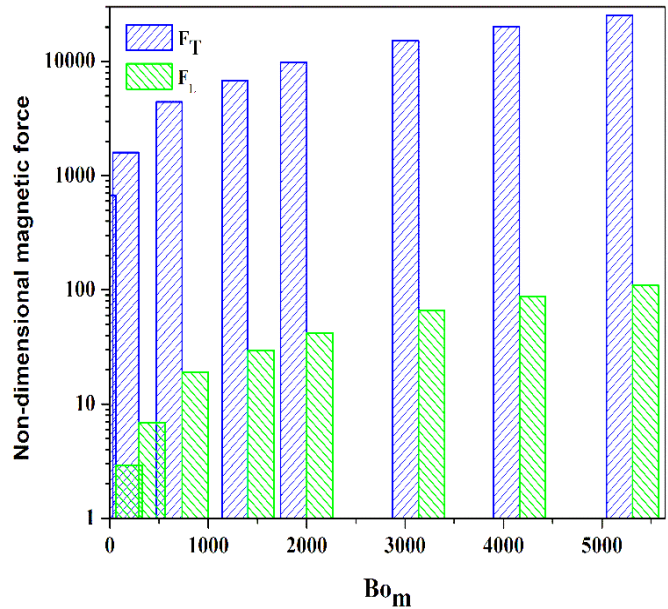

**Fig.6:** Variation of the non-dimensional magnetic force with $Bo_m$. $F_T$ and $F_L$ represent the magnetic forces along the transverse and longitudinal directions, respectively. The magnetic forces are normalized by the initial interfacial force ($F_S = \pi D_o \sigma_{lv}$).

At higher $Bo_m$ (~5310), holes are nucleated at different locations of the liquid lamella, which is similar to rupture of thin films upon impact [21, 32] (fig.5 (a) and (c), shown by arrows). For a fixed We=60, the lamella (at the maximum spread state) was observed to become highly unstable due to the onset of rupturing within the film, and the successive proliferation with increasing $Bo_m$. This type of internal rupture of fluid films has been observed earlier at high We regimes (We ~ 800-7200) for droplet impact dynamics without any magnetic field [20]. Rupture of films in absence of magnetic field is only possible when the surface energy of the lamella-substrate system becomes equal to that of hole-liquid-substrate system. This can be created by the high inertial energy of high We regime impact. In the present study, we showed that due to ferrohydrodynamic forces within the spreading droplet in presence of magnetic field, film rupture is triggered at We as low as ~60 (at least an order of magnitude lower than the cases without magnetic field).

Further, we have developed an analytical model along the lines of the theory for the change in energy [21] of the lamella with and without the hole(s). At higher We and no magnetic field, the liquid lamella is ruptured by air bubbles entrapped underneath the impinging droplet or due to the surface roughness, through nucleation of hole(s) and their growth. We observed that at a fixed We, the thickness of the liquid lamella decreases with increase in $Bo_m$. The liquid lamella is punctured at different locations at higher $Bo_m$, as shown by the red arrow marks in fig.5 (a) and (c), indicating that the nucleation of the holes is triggered by the ferrohydrodynamic forces.

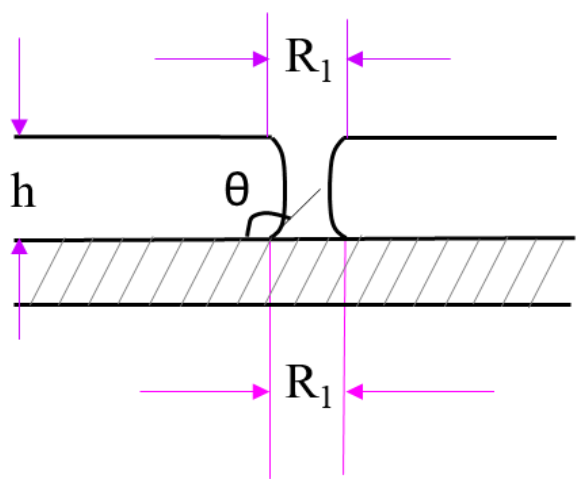

**Fig.7** Schematic of the hole profile at a certain contact angle (θ).

First, we assume that the radius of all the nucleated holes is equal at the liquid-solid interface and at the liquid-air interface, as illustrated in fig.7. In the presence of horizontal magnetic field, the surface energy of the liquid lamella can be deduced by considering the effective diameter of the lamella after impact. The surface energy of the liquid film without hole nucleation can be expressed as

$$E_{film} = (\sigma_{lv} + \sigma_{sl}) A_a \quad (14)$$

Where, $\sigma_{sv}$ is the surface energy of the solid-air interface, and $A_a$ is the average area of the film before formation of hole(s).

The average area $A_a$ is considered to account for the asymmetric orthogonal spreading of the droplet after impact when the droplet attains maximum spreading diameter. Since the fluid film is punctured with nucleated holes of radius $R_1$ (see fig. 7) at $Bo_m > 300$, the surface energy of the film with hole can be expressed as

$$E_{hole} = \sigma_{lv}(A_a - \pi R_1^2) + \sigma_{sl}(A_a - \pi R_1^2) + S\sigma_{lv} + \pi R_1^2 \sigma_{sv} \quad (15)$$

$$\text{where } S = \frac{1}{2}\pi R_1^2 \left\{ (1+\cos^2\theta)*\sinh\left(2\frac{h/R_1}{\sin\theta}\right) + 2\cos\theta\left[\cosh\left(2\frac{h/R_1}{\sin\theta}\right) - 1\right] + 2\frac{h}{R_1}\sin\theta \right\}$$

The surface area (S) of the liquid meniscus is determined from the Young-Laplace equation of capillarity [32]. The change in surface energy of the liquid lamella-hole system is expressible as

$$\Delta E = \sigma_{lv}[S - \pi R_1^2(1-\cos\theta)] \quad (16)$$

The rupture of the liquid lamella depends upon the magnitude of $\Delta E$. The value of S depends upon the magnitude of the solid–liquid contact angle and the ratio ($h/R_1$).

The nucleation of the holes and subsequent proliferation towards destabilizing the lamella is triggered for the case $\Delta E < 0$. At inception, the area of the holes is very small in comparison to the whole lamella. From eqns. (15) and (16), this physically translates to $\theta \sim 90°$. Similarly, when $\theta$ approaches $0°$ (a completely spread out lamella), the magnitude of $\Delta E$ is always positive. This is of course possible only for wetting surfaces, which is not the present case on SH surfaces. On the contrary, when $\Delta E < 0$, the hole(s) reduce the net energy of the lamella. Thus, proliferation and growth of the holes during retraction phase and as it is a thermodynamically favorable state for the spread out lamella. Equating $\Delta E = 0$, we obtain the critical thickness ($h_{critical}$) at which the ferrohydrodynamic interactions lead to hole nucleation and rupture as

$$\sinh\left[\frac{2h_{critical}}{R_1}\frac{1}{\sin\theta}\right] + \cos^2\theta \sinh\left[\frac{2h_{critical}}{R_1}\frac{1}{\sin\theta}\right] + 2\cos\theta\cosh\left[\frac{2h_{critical}}{R_1}\frac{1}{\sin\theta}\right] + \frac{2h_{critical}}{R_1}\sin\theta = 2 \quad (17)$$

The critical thickness ($h_{critical}$) obtained from the energy analysis can be correlated to the experimentally noted thickness of liquid lamella after impact, at the maximum spread state.

We derive this thickness through mass conservation of the droplet before impact and at maximum spread. This can be expressed as

$$\frac{4}{3}\rho\pi R_o^3 = \rho\pi R_{max}^2 h \quad (18)$$

Eq. (18) can be simplified

$$h = \frac{4R_o}{3\psi_m^2} \quad (19)$$

We now focus towards developing a relationship between the orthogonal spreading ratio ($\xi$), the average maximum spreading factor, and the ratio ($Bo_m/We_o$) from the experimental observations. This is essential as the $h_{critical}$ cannot be measured accurately from image processing and hence needs to be represented in terms of the spreading parameters. As can be seen from fig.8 (a), $\xi$ and $\psi_{m,average}$ conform to an approximately linear trend with respect to the quantity $\left[\frac{Bo_m}{We}\right]^{0.5}$. The prediction curves using this function form approximates the experimental data within 5% error. Based on this observation, we scale the maximum spreading factor as $\psi_m \approx \left[\frac{Bo_m}{We}\right]^{0.5}$. Eqn. (19) is now expressible as

$$h \approx \frac{4R_o}{3} \frac{We}{Bo_m} \quad (20)$$

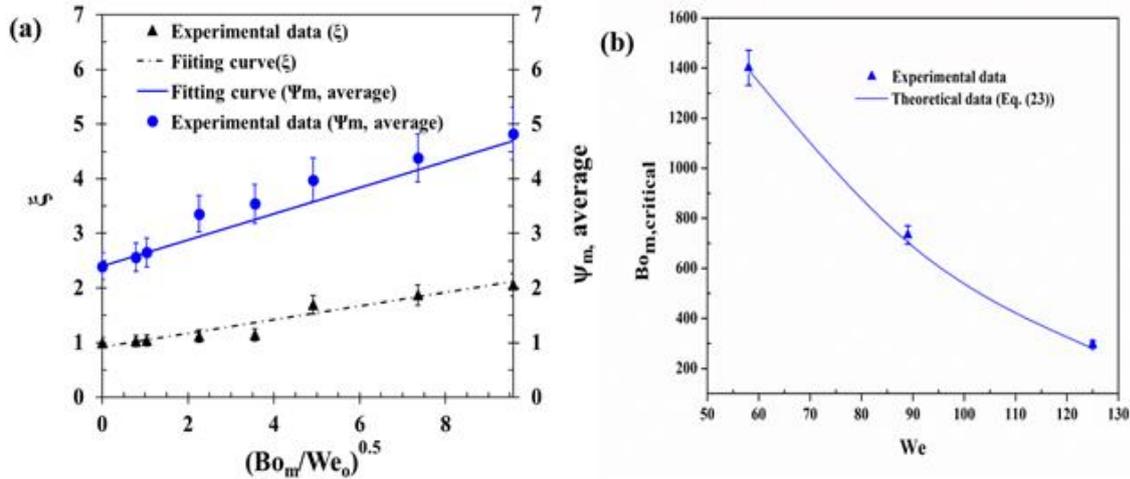

**Fig.8 (a)** The average maximum spreading factor (left right axis) and orthogonal spreading ratio (left y axis) scaled against $(Bo_m/We)^{1/2}$. $\psi_{m,\ average}$ represents the average maximum spreading factor and it is estimated by considering the geometric mean of maximum spreading factor along the longitudinal and transverse direction of the applied magnetic field (b) Comparison of theoretical and experimental critical $Bo_m$. It is observed that with increasing We, the rupturing instability sets at lower critical $Bo_m$. This indicates the role of ferrohydrodynamics towards the onset of the instability.

Eqns. (17) and (20) can be combined and written in non-dimensional form as

$$\sinh\left[\frac{2h''}{R_1''}\frac{1}{\sin\theta}\right] + \cos^2\theta \sinh\left[\frac{2h''}{R_1''}\frac{1}{\sin\theta}\right] + 2\cos\theta\cosh\left[\frac{2h''}{R_1''}\frac{1}{\sin\theta}\right] + \frac{2h''}{R_1''\sin\theta} = 2 \qquad (21)$$

Where, $h' = \frac{2}{3}\frac{We}{Bo_m}$ (22)

And $h'' = \frac{h_{critical}}{D_o}$, $h' = \frac{h}{D_o}$ and $R_1' = \frac{R_1}{D_o}$. For $h'' = h'$, and substituting Eq. (22) into Eq. (21), the condition for inception of holes and subsequent rupture of the liquid lamella due to ferrohydrodynamic interactions is expressed as:

$$\sinh\left[\frac{4We}{3Bo_{m,critical}R_1''}\frac{1}{\sin\theta_r}\right] + \cos^2\theta_r \sinh\left[\frac{4We}{3Bo_{m,critical}R_1''}\frac{1}{\sin\theta_r}\right] + 2\cos\theta_r\cosh\left[\frac{4We}{3Bo_{m,critical}R_1''}\frac{1}{\sin\theta_r}\right] + \frac{4We}{3Bo_{m,critical}R_1''}\sin\theta_r = 2 \qquad (23)$$

where $Bo_{m,critical}$ is the critical magnetic Bond number, at which the rupture of liquid lamella occurs through proliferation of the nucleated holes. For a fixed We, $Bo_{m,critical}$ is a function of the solid-liquid contact angle, the radius of holes, and the Ha of the ferrofluid droplet. Fig.8b compares the theoretical critical $Bo_m$ with the experimental observations at which the hole nucleation and proliferation across the film after impact upon SH surface is visually evident. The experiments were performed in a controlled manner at fixed We, and gradually increasing $Bo_m$, to identify the critical value at which the lamella begins to exhibit hole nucleation and rupture via the ferrohydrodynamic instability. Previous studies [21, 34] reported the dependence of the growth rate of hole(s) in the liquid lamella on $\theta_r$. In the current study, we consider the receding contact angle in the model as $\theta_r \sim 135.5^o$. The radius of the hole(s) is noted to vary from ~80 μm to ~850 μm after impact. In the present study, there was no nucleation of holes even at We=120 in the absence of magnetic field. However, for a critical $Bo_m$, the rupture instability occurs even at low We=60. Hence, it can be concluded that the rupture instability of the liquid lamella is triggered by the ferrohydrodynamic interactivities within the spreading-retracting droplet.

Finally, we show that the ferrohydrodynamic instabilities noted in the present case are ubiquitous to SH surface only. In order to verify the influence of wettability, we have performed the studies on hydrophobic Teflon and hydrophilic glass surface. At We=60 and $Bo_m$=5310 no hole(s) formation was observed in the liquid lamella during droplet retraction on the glass surface (see fig. 9). Our experimental data are in agreement with reports on impact study upon glass [18], where the stable liquid lamella does not show any nucleation of holes after impact even at We=800. Previous studies [20, 33] suggested that the onset of rupture of liquid lamella depends on the solid-liquid contact angle. The liquid lamella ruptures at very high We=7200 [18] if the contact angles are either very small or very large. In the present case, nucleation of hole(s) is absent at We=60 for Teflon (the equilibrium contact angle: $\theta_e=87^o$) and glass ($\theta_e=37^o$) in presence of magnetic field ($Bo_m$=5310), whereas the magnetic field triggers shattering of the liquid lamella at the same We and $Bo_m$ on the SH surface ($\theta_e=145.6^o$).

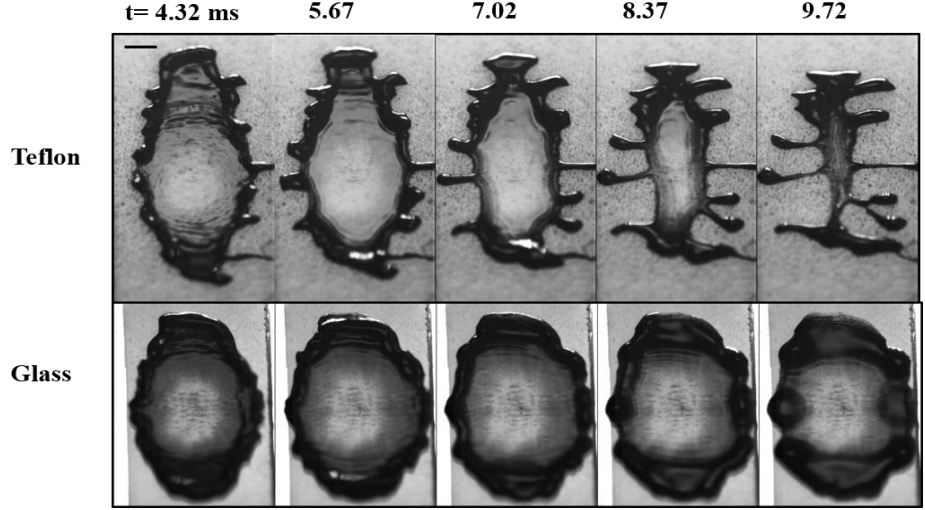

**Fig.9**: Maximum spreading and retraction regime images of low Ha ferrofluid droplets on variant wettability surface for a fixed We =60 and $Bo_m$=5310. The scale bar is 6 mm.

As closure, we propose a phase map (fig. 10) of post-impact droplet ferrohydrodynamics for a range of Weber (10<We<130) and magnetic Bond numbers (0<$Bo_m$< 6000). Similarly, the range of $Ca_m$ (0.3<$Ca_m$<30) and Ha (0<Ha<15) are also considered corresponding to the aforesaid We and $Bo_m$ range to develop the phase map. The magnetic capillary number ($Ca_m = \dfrac{\eta V_o}{\sigma_{lv}}$) is expressed as the ratio of viscous force to surface tension force. It is observed that experiments are not possible below We=40 and $Bo_m$>10, since the needle of the syringe is present within the magnetic field, and distorts the formation and detachment of the droplet. The various phenomena noted for different Weber number and Bond number are as follows:

(a) At 10<We<125 and $Bo_m$<1 (Ha<0.01 and 0.3<$Ca_m$<30), we observed three major outcomes after an impact, viz. complete rebound, rebound with pinch off, and crown filament structure (see fig.10). In this regime, the effect of magnetic force is absent and there is only competition between the inertial and interfacial forces along with viscous dissipation. This wide range of post-impact droplet dynamics is earlier reported [13] in absence of magnetic field, by varying the impact velocity ($V_o$=0.5-6.0 m/s) and viscosity ($\eta$ =1-100 mPa-s).

(b) The complete rebound occurs at 40≥We<60 and 25<$Bo_m$<300. In this regime, there is a balance between inertial and magnetic forces, irrespective of the viscous dissipation. Again, the post-impact droplet shows rebound suppression due to initiation of asymmetric spreading in orthogonal direction at 40≥We≤60 and $Bo_m$~300 as shown in Fig.10a and b via open pentagon.

(c) Similarly, at 40<We<125 and 300<$Bo_m$<6000, the shattering of liquid lamella always take place since the magnetic force is dominant over the inertial force of pre-impact drops (see fig.10a and b). It is also observed that the lamella ruptures at low $Bo_m$ range (10<$Bo_m$<300) and higher We =125. With high impact velocity, the rim is under the influence of higher inertial forces, leading to lamella breakup even at lower critical $Bo_m$ regimes.

(d) Though the present study did not conduct any detailed experiments beyond $We > 125$ and $300 < Bo_m < 6000$, for the few cases tested, the shattering of lamella always occurs as both inertial and magnetic forces are dominant over surface tension and viscous forces.

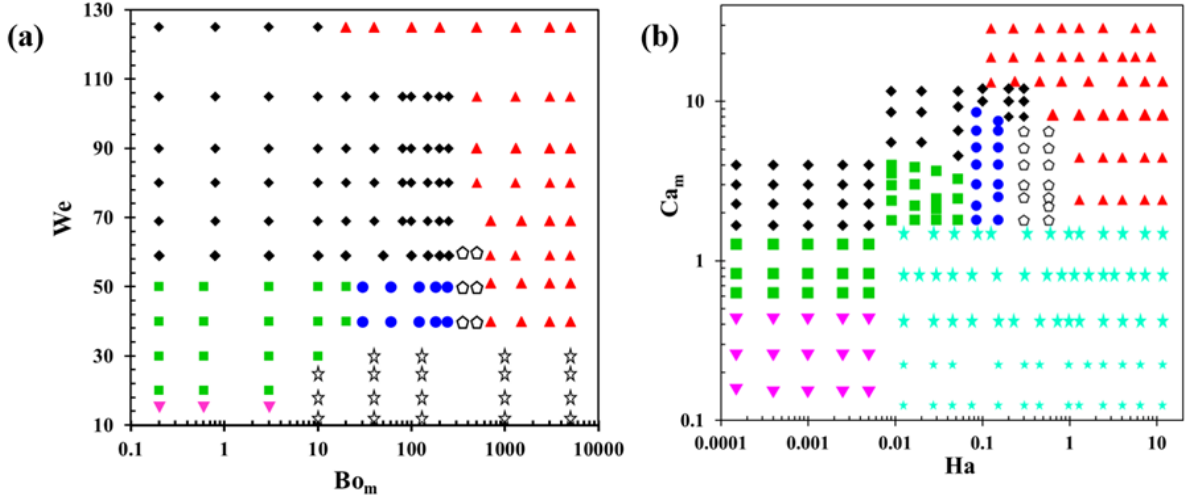

**Fig.10**: Regime maps of post-impact ferrofluid drops upon SH surface with consideration of different non-dimensional numbers. Pink inverted triangle (▼) and red triangle (▲) represent complete rebound and shattering of liquid lamella (with ferrohydrodynamic instabilities), respectively, both in absence of magnetic field. Green Square (■) and black rhombus (◆) represent complete rebound with pinch off and crown structure formation. Cyan pentagon (⬠) and star (☆) represent magnetic rebound suppression and no experimental region. Blue circle (●) represents reduced rebound (with arrested pinch-off) in presence of magnetic field. The value of $Ha < 0.01$ is considered as the droplet impact in absence of magnetic field.

## 4. Conclusions

In summary, we have investigated the impact dynamics of ferrofluid drops through experiments and theoretical models under the influence of horizontal magnetic field. The rebound suppression is significantly observed impacting upon SH surface at low $Bo_m \sim 300$. We analytically predict that the rebound behavior is subdued by non-uniform distribution of kinetic energy along the longitudinal and transverse direction of the magnetic field lines. The orthogonal spreading manifests through variant magnitude of Lorentz force acting on the post impact droplet.

Further, with increase in $Bo_m$ ($Bo_m > 300$) we found that during the retraction phase, the liquid lamella becomes unstable through proliferation of holes after attaining a critical thickness. The instabilities prompted by ferrohydrodynamic interaction ruptures the liquid lamella upon SH surface. In case of hydrophilic glass and Teflon, such instabilities are even absent at the higher value of the $Bo_m$ ($Bo_m \sim 5310$), compared to that of SH surface case. Our analytical model approximates the critical $Bo_m$ for a fixed We. Finally, the phase diagram encompassing the impact phenomena under a wide range of We and $Bo_m$ was presented.

## Acknowledgements

NS and GK would like to thank the Ministry of Human Resource Development, Govt. of India, for the doctoral scholarships. DS and PD would like to thank IIT Ropar for funding the present work (vide grants 9-246/2016/IITRPR/144 & IITRPR/Research/193 respectively). PD also thanks IIT Kharagpur for partially funding the work.

## Conflicts of interests
The authors do not have any conflicts of interests with any individuals or agencies with respect to the current research work.